\documentclass{aip-cp}

\usepackage[numbers]{natbib}
\usepackage{rotating}
\usepackage{graphicx,accents}

\def\bea{\begin{eqnarray}}
\def\eea{\end{eqnarray}}
\def\sea{\nonumber \\ &&}
\def\lla{\left\langle}
\def\rra{\right\rangle}
\def\za{\alpha}
\def\zb{\beta}

\def\ssc{\scriptscriptstyle}
\def\lsim{\mathrel{\raise.3ex\hbox{$<$\kern-.75em\lower1ex\hbox{$\sim$}}} }
\def\gsim{\mathrel{\raise.3ex\hbox{$>$\kern-.75em\lower1ex\hbox{$\sim$}}} }

\makeatletter
\DeclareRobustCommand{\cev}[1]{%
  \mathpalette\do@cev{#1}%
}
\newcommand{\do@cev}[2]{%
  \fix@cev{#1}{+}%
  \reflectbox{$\m@th#1\vec{\reflectbox{$\fix@cev{#1}{-}\m@th#1#2\fix@cev{#1}{+}$}}$}%
  \fix@cev{#1}{-}%
}
\newcommand{\fix@cev}[2]{%
  \ifx#1\displaystyle
    \mkern#23mu
  \else
    \ifx#1\textstyle
      \mkern#23mu
    \else
      \ifx#1\scriptstyle
        \mkern#22mu
      \else
        \mkern#22mu
      \fi
    \fi
  \fi
}

\makeatother

\begin{document}

\thispagestyle{empty}
\begin{flushright}
NCU-HEP-k074  \\
Aug 2018
\end{flushright}

\vspace*{.5in}

\begin{center}
{\bf Quantum Spacetime Pictures and Dynamics from a Relativity Perspective}
\\
\vspace*{.5in}
{\textbf  Otto C.W. Kong}\\[.05in]
{\it Department of Physics and 
Center for High Energy and High Field Physics,\\
National Central University, Chung-Li 32054, Taiwan\\
E-mail: otto@phy.ncu.edu.tw}
\vspace*{1.in}
\end{center}
{\textbf Abstract :}\ \
Based on an identified quantum relativity symmetry the contraction of 
which gives the Newtonian approximation of Galilean relativity, a 
quantum model of the physical space can be formulated with the Newtonian 
space seen in a way as the classical approximation. Matching picture for 
the observable algebra as the corresponding representation of the
group $C*$-algebra, describes the full dynamical pictures equally 
successfully. Extension of the scheme to a Lorentz covariant setting 
and beyond will also be addressed. The formulation of quantum 
mechanics allows the theory to be seen in a new picture in
line with the notion of a noncommutative spacetime.


\vfill
\noindent --------------- \\
$^\star$ Talk presented  at the 
 10th Jubilee International Conference of the Balkan Physical Union,
Aug 26-30, 2018, Sofia, Bulgaria.

\clearpage
\addtocounter{page}{-1}


\title{Quantum Spacetime Pictures and Dynamics from a Relativity Perspective}

\author[aff1]{Otto C.W. Kong
}
\eaddress{otto@phy.ncu.edu.tw}

\affil[aff1]{Department of Physics and 
Center for High Energy and High Field Physics,\\
National Central University, Chung-Li 32054, Taiwan}

\maketitle

\begin{abstract}
Based on an identified quantum relativity symmetry the contraction of 
which gives the Newtonian approximation of Galilean relativity, a 
quantum model of the physical space can be formulated with the Newtonian 
space seen in a way as the classical approximation. Matching picture for 
the observable algebra as the corresponding representation of the
group $C*$-algebra, describes the full dynamical pictures equally 
successfully. Extension of the scheme to a Lorentz covariant setting 
and beyond will also be addressed. The formulation of quantum 
mechanics allows the theory to be seen in a new picture in
line with the notion of a noncommutative spacetime.
\end{abstract}

\section{INTRODUCTION}
In physics, experiments ultimately inform us as to what constitutes a good 
theoretical model of any physical concept: the physical space should be no 
exception. In a theory of particle dynamics like the Newtonian one, there 
is no physical picture of the physical space itself beyond that of the 
configuration space of a free particle (or the center of mass of a closed 
system of particles). The physical space is simply the totality of all possible
positions for a physical object. Hence, the model for the physical space
is not to be taken as independent of the physical theory. For a theory of
particle dynamics, in particular, the model for the physical space assumed
is only as good as the theory itself. We have seen how Einstein's theory
of special relativity illustrate that the Newtonian model of the physical
space is only to be taken as an inseparable part of the Minkowski spacetime.
Quantum mechanics as it is to date, however, inherits with little critical 
revision many Newtonian conceptual notions. Most importantly, the
Newtonian space model is still assumed, even though we are then forced
to conclude that a physical object has no definite position. Our key
question is if there is actually a different, better, model of the physical
space for quantum mechanics.  We want to see how the theory informs 
us as to what space is like, or what is the quantum model for it and how 
that is related to the more familiar Newtonian picture, which one must 
be able to retrieve as a limit or an approximation.

Quantum mechanics has position observables described by operators
which are parts of the basic ingredients of an noncommutative algebra 
of observables. If the positions are modeled by a mathematical
structure different from that of classical commutative observable
algebra, the physical space they depict should be given by a different
mathematical model. That naive thinking is behind efforts to 
construct quantum/noncommutative geometric models of the
spacetime which has become quite a endeavor \cite{nsp,nsp2,N-s}.
However, such efforts miss the true well established physical theory
of the kind, namely simple quantum mechanics. Well, the position
operators in quantum mechanics commutate among themselves.
The latter seems to suggest that the quantum configuration space
is like classical or commutative, only the quantum phase space is 
not. That perhaps is where the difficulty has been. In fact, we have
a clear notion of the quantum phase space for a free particle as
different from the classical one. It is given by the infinite
dimensional (projective) Hilbert space. But there is no notion of
the quantum configuration space. The main results, from our
recent works \cite{066,070}, we will sketch here is a full answer
to all that. In a single statement, the project Hilbert space is the
proper model for the physical space as behind quantum mechanics.
As the space of pure states of the observable algebra, it is a
mathematical structure dual to the latter \cite{dual} and hence
can be seen as an alternative description of the noncommutative
geometry. The configuration part and the momentum part of it are 
like the space and time parts of the Minkowski spacetime, they are 
not separable notions and can only truly be described independently
in the Newtonian approximation. The Newtonian (configuration)
space and phase space can be retrieved as approximations of
the quantum model. As said, the model of the physical space
is an integral part of the dynamical theory. Our analysis gives
the complete relation between whole dynamics theories. 
Quantum mechanics is really a theory of particle dynamics on
the quantum physical space, the classical approximation of
which is Newtonian mechanics on the Newtonian space.  

The configuration space and the phase space can be constructed 
as a representation space for the relativity symmetry. We can
start from the Poincar\'e symmetry of special relativity with
coset spaces $ISO(1,3)/SO(3) \times T$ ($T$ denotes the time
translation subgroup) and $ISO(1,3)/SO(1,3)$ giving the
phase space for an Einstein particle and the Minkowski
spacetime, respectively. The Newtonian approximation is to
be retrieved as $c \to \infty$ which is best formulated as
contraction limit \cite{060,G} of the relativity symmetry
$ISO(1,3) \to  G(3)$, the Galilean group \cite{067}. For the case
of quantum mechanics, one should take the $U(1)$ central 
extension of the Galilean group $\tilde{G}(3)$ as the true 
relativity symmetry. A contraction as like $\hbar \to 0$ limit 
trivializes the central extension and retrieves the classical results. 

Our background framework is one of deformations of
special relativity \cite{030,dsr,dsr2}. Symmetry deformation is
essentially the inverse of symmetry contraction. The latter
can be much more rigorously implemented to all aspects of
the theory formulated from the symmetry. We have in mind
an ultimate quantum relativity symmetry that is kind of fully
noncommutative and stable against further deformation
and having essentially all fundamental constants $G$, $\hbar$,
and $c$ incorporated as like structural constants of the Lie
algebra. The whole mathematical scheme as we illustrate
for the full dynamics quantum to classical can be performed
for the candidate theories at any level of the relativity
symmetries as some contraction of the original. A specific
contraction exactly gives a theory as an approximation to
the one at the higher level. In fact, we believe to any 
dynamical theory of noncommutative/quantum spacetime
geometry, our results may serve as the crucial first link from
the bottom-up.  

In the next section, we will give a sketch of the scheme
of mathematical formulation which is expected to work
in general. After that we will present a bit of details for the 
explicit case of quantum mechanics and its Newtonian
approximation. 

\section{SCHEME OF MATHEMATICAL FORMULATION}
A plausible picture of the full relativity symmetry contraction
scheme can be given as 
\bea 
{SO(2,4)}  \quad\longrightarrow   && {H_{\!\ssc R}(1,3)}
                      \quad\qquad \longrightarrow \quad {H_{\!\ssc G\!H}(3)} \supset \tilde{G}(3)  \supset H_{\!\ssc R}(3)    \supset H_(3)
\sea
 \;\;\; \downarrow   \hspace*{1.2in} \qquad \quad\downarrow  
\sea \hspace*{-.6in}
 ISO(1,3) \subset S(1,3)
                      \quad\qquad \longrightarrow \quad
S_{\!\ssc G}(3) \supset {G}(3) 
\nonumber
\eea
The $SO(2,4)$ symmetry has the Lorentz symmetry $SO(1,3)$
plus nine generators which can be seen as two four-vectors
for essentially $X_\mu$ and $P_\mu$ and their commutator
of the Heisenberg form. Commutators among the $X_\mu$
set, and also among the $P_\mu$ set, are $SO(1,3)$ generators.
The limit  the two four-vector sets each has mutually commutative 
components is the ${H_{\!\ssc R}(1,3)}$ symmetry, a Heisenberg-Weyl 
group with the $1+3$ sets of  $X_\mu$ and $P_\mu$ supplemented
by the rotation symmetry within each set. $S(1,3)$ is an 
extension of the Poincar\'e symmetry and is the classical limit of
${H_{\!\ssc R}(1,3)}$ with the Heisenberg commutators trivialized.
Each arrow can be implemented as a contraction. The horizontal
arrows apart from the first one is the $c \to \infty$ ones. Details
of the contractions together with some discussion of the physics
pictures as seen from the relevant coset spaces are presented
in Ref.\cite{071}, illustrating a story consistent with all known
physics. The ${H_{\!\ssc R}(1,3)}$ would give a Lorentz covariant
quantum mechanics. Under the framework, the relativity 
symmetry for quantum mechanics has to be extended to the
${H_{\!\ssc G\!H}(3)}$ group and for Newtonian mechanics
the $S_{\!\ssc G}(3)$ group. For the relevant representation
of focus here, the extensions do not matter. We have only to
focus on an irreducible representation of the $H(3)$ subgroup
for the quantum cases and its classical limit. The rest of the
group acts on the representation in an essentially trivial
manner. For example, such a representation is the 
familiar spin zero time independent representation of
the $\tilde{G}(3)$ group. Note that technically, we use $X_i$ 
in the place of the generators for Galilean boosts  which 
should be taken as $mX_i$ for a particle of mass $m$.

With a Lie group serving as the relativity symmetry, the model 
for the physical space is to be given by a representation. An
irreducible unitary representation is naturally served as a
cyclic topological irreducible $*$-representation of the group
$C^*$-algebra which serve as the algebra of observables. Guided
by the case of quantum mechanics, the project Hilbert space for
such a representation may is the space of pure states for the
algebra, obtainable from the GNS construction \cite{Da}. From
the group theoretical point of view, the unitary representation
can be formulated as a coherent state representation with
each coherent state corresponds to point(s) on a coset space
which may not be a vector space in general. Such spaces of pure 
state are typically K\"ahler manifolds with a symplectic 
structure. One has a natural formulation of Hamiltonian 
dynamics on the phase space which corresponds exactly to
Schr\"odinger dynamics on the Hilbert space \cite{CMP}. There 
is a Heisenberg picture of the dynamics as automorphism flows
on the observable algebra.  To implement a contraction, the
effect of the contraction should be trace through the specific
representation. The limiting results for the model of the
physical space as well as the full dynamical theory can be 
retrieved. 

Let us have a few more words on how the scheme works for
the classical approximation to a quantum theory. At the 
classical level, we have a commutative observable algebra,
irreducible representations of which has to be one dimensional.
The quantum observable algebra has to contract to a 
commutative limit. The Hilbert space then reduces to the
individual one dimensional subspaces each corresponds
exactly to one pure state. Superposition principle is lost
as superposition of two classical pure states is only a mixed
state.  The Hilbert space picture is one of Koopman-von
Neumann \cite{KvN}. The `projective Hilbert space' as the
space of pure states reduces to the classical coset space; the
canonical coherent states are essentially the classical states
and the only ones that survive the contraction as pure states.
The Heisenberg dynamics on the observable algebra reduces 
to Poisson dynamics. The latter is a familiar fact which is
behind the idea of deformation quantization. From our
relativity symmetry perspective, hence, the deformation
in deformation quantization is really the deformation of
the representation of the group $C^*$-algebra as a
consequence of the deformation of the relativity symmetry
group and the corresponding representation.

\section{THE PROJECTIVE HILBERT SPACE MODELS}
We write quantum expressions with the $\hbar=2$ units here. This 
choice gives the Heisenberg commutation algebra the `unusual' form of 
\bea
[X_i, P_j]=2i\delta_{ij} I \;,
\eea
while setting the minimal uncertainty product to unity. 
The  (canonical) coherent state representation is given by
\bea
\left|p^i,x^i\rra = {U}(p^i,x^i) \left|0\rra \equiv
e^{-i\theta} U(p^i,x^i,\theta) \left|0\rra
\eea
where
\bea
U(p^i,x^i,\theta) \equiv 
e^{ix_i p^i \hat{I}}
e^{i\theta \hat{I}} 
e^{-ix^i\hat{P}_i} 
e^{ip^i\hat{X}_i} 
= e^{i(p^i\hat{X}_i- x^i\hat{P}_i +\theta\hat{I})} \;,
\eea
$\left|0\rra \equiv \left|0,0\rra$ is a fiducial normalized vector, 
$\hat{X}_i$ and $\hat{P}_i$ are representations of the generators $X_i$ 
and $P_i$ as self-adjoint operators on the abstract Hilbert space  $\mathcal{H}$ 
spanned by the $\left|p^i,x^i\rra$ vectors, and 
$\hat{I}$ is the identity operator representing the central generator $I$. 
 $(p^i,x^i,\theta)$ corresponds to a generic element of the 
Heisenberg-Weyl $H(3)$ group as  
 \bea
{W}(p^i,x^i,\theta) = \exp [i(p^iX_i- x^iP_i +\theta I)],
\eea
with
\bea\label{ww}
{W}(p'^i, x'^i, \theta')  {W}(p^i, x^i, \theta)
= {W}\!\left(p'^i+p^i, x'^i+x^i, \theta'+\theta-(x'_i p^i - p'_i x^i) \right) \;.
\eea
Note that we have $p^i$ and $x^i$ here corresponding to {\em half} the 
expectation values of $\hat{P}_i$  and $\hat{X}_i$. On the wavefunctions 
$\phi (p,x )\equiv \lla p,x| \phi \rra$, we have (suppressing the $i$-index)
\bea 
 \lla p,x \left| \hat{X} \right|\phi \rra &=& \hat{X}^{\!\ssc L}  \phi (p,x ) \;,
\nonumber \\
\lla p,x \left| \hat{P} \right|\phi \rra &=& \hat{P}^{\!\ssc L}  \phi (p,x ) \;,
\eea
where
\bea
\hat{X}^{\!\ssc L} &=&  x +  i \partial_{p}\;,
\nonumber \\
\hat{P}^{\!\ssc L} &=& p -  i \partial_{x}\;,
\label{L}
\eea
and
\bea \label{u-shift}
{U}^{\ssc\! L}(p,x) \phi (p',x') \equiv
 \lla p',x' \left|{U}(p,x)  \right|\phi \rra 
= \phi (p'-p,x'-x ) e^{{i}(px'-xp')}\;.
\eea
We can see again that 
$\hat{P}^{\!\ssc L}$ and $\hat{X}^{\!\ssc L}$ generate translations in
$x$ and $p$, respectively. The wavefunction $\phi_a (p,x)$ 
for the coherent state $\left|p_a,x_a \rra$  is given by 
\bea \label{csg}
\phi_a (p,x) \equiv  \lla p,x |p_a,x_a \rra 
= e^{i(p_ax-x_ap)} e^{-\frac{1}{2}[(p-p_a)^2+(x-x_a)^2]} \;;
\eea
in particular, the $\left| 0,0\rra$ state wavefunction is denoted by
$\phi_0 (p,x)$ and $\phi_0 (p,x)= e^{-\frac{1}{2}(p^ip_i+x^ix_i)}$, which is  
a symmetric Gaussian of unit width. The expression $ \lla p,x |p_a,x_a \rra$ 
may also be taken as giving the overlap of two different coherent states. 
We denote the Hilbert space  of wavefunctions by $\mathcal{K}$, on which 
we have  $\frac{1}{\pi^n}  \int\!\! dp dx \bar{\phi}(p,x){\phi'}(p,x)
 =\langle {\phi}| \hat{I} |\phi'\rangle$ with
 $\hat{I}=\frac{1}{\pi^n} \int dp dx \left|p,x \rra \!\!\lla p,x \right|$,
which keeps $\phi_a (p,x)$  as a normalized wavefunction ($n=3$).   
Apart from a set of measure zero, all irreducible representations of
$H(3)$ are essentially the same, up to rescaling of 
$\hat{P}^{\!\ssc L}$ and $\hat{X}^{\!\ssc L}$ which can be taken
as a different choice of the physical unit to measure them. 

The contraction to give the classical approximation is given by
$X_i^c=\frac{1}{k} X_i$ and  $P_i^c=\frac{1}{k} P_i$ with {$k \to \infty$},
hence $[X_i^c,P_i^c] \to 0$.
With the state $\left| p_i,x_i \rra$ re-labeled as $\left| p^c_i,x^c_i \rra$ 
by the corresponding classical values, namely $p^c_i =  \frac{1}{k}  p_i$
and $x^c_i =  \frac{1}{k}  x_i$ taken as finite, we have
\bea
\lla {p}'^c_{i},{x}'^c_{i} \right| \hat{X}_i^c \left|{p}_i^c,{x}_i^c\rra
&=& [({x}'^c_{i}+{x}^c_{i})-i({p}'^c_{i}-{p}^c_{i})]
{\lla {p}'^c_{i},{x}'^c_{i} |{p}_i^c,{x}_i^c\rra } 
\nonumber \\
\lla {p}'^c_{i},{x}'^c_{i}\right| \hat{P}_i^c \left|{p}_i^c,{x}_i^c\rra
&=& [({p}'^c_{i}+{p}^c_{i})+i({x}'^c_{i}-{x}^c_{i})]
{\lla {p}'^c_{i},{x}'^c_{i} |{p}_i^c,{x}_i^c\rra} 
\eea
with
\bea
{\lla {p}'^c_{i},{x}'^c_{i} |{p}_i^c,{x}_i^c\rra} =
\exp \left[ik^2 ({{x}'^c_{i} {p}^c_{i} - {p}'^c_{i} {x}^c_{i}})\right] 
{\exp\left[-\frac{k^2}{2}[({x}'^c-{x}^c)^2+({p}'^c-{p}^c)^2]\right]} \to 0
\eea
and
 ${\lla {p}^c_{i},{x}^c_{i} |{p}_i^c,{x}_i^c\rra} =1$. Hence, the overlap
between any two coherent states vanishes and the $\hat{X}_i^c$
and  $\hat{P}_i^c$ are completely diagonal. Each  
$\left| p^c_i,x^c_i \rra$ gives a one dimensional irreducible
representation of the contracted symmetry. We will see that as a
representation for the group $C^*$-algebra for the full relativity
symmetry group, the observable algebra is given by functions
of  $\hat{X}_i$ and  $\hat{P}_i$ and hence functions of $\hat{X}_i^c$
and  $\hat{P}_i^c$ at the classical limit. In the language of 
$\mathcal{K}$, the coherent state wavefunction of 
Equation~(\ref{csg}) goes to a delta function centered on 
$(p^c_a, x^c_a)$, with 
$\hat{X}^{c\ssc L} = x^c +\frac{i}{k^2}\partial_{p^c} \to x^c$
and 
$\hat{P}^{c\ssc L} = p^c -\frac{i}{k^2}\partial_{x^c} \to p^c$.

The $H(3)$ is isomorphic as a manifold to the coset space
$\tilde{G}(3)/SO(3)\times T$. We illustrate also the contraction 
picture of the coset space as
\bea \label{psc}
\left(\begin{array}{c}
dp^i_c  \\ dx^i_c \\ d\theta \\    0
\end{array}\right) =
\left(\begin{array}{cccc}
 \omega^i_j &  0& 0  &  \bar{p}^i_c \\
0  & \omega^i_j &  0 &  \bar{x}^i_c  \\
-{\frac{1}{k^2}}\bar{x}_{cj}   & {\frac{1}{k^2}}\bar{p}_{cj}   &  0 & \bar{\theta} \\
 {0}   & {0} &  0 & 0
\end{array}\right)
\left(\begin{array}{c}
p^j_c  \\ x^j_c   \\ \theta \\    1
\end{array}\right) \;;
\eea
note that here $x_c$ and $p_c$ are group parameters
satisfying $x_c = k x$ and $p_c = k p$ which are different
from the coherent state labels $x^c$ and $p^c$ 
($p_c \hat{X}^c - x_c \hat{P}^c = p \hat{X} -x \hat{P}$). 
At the classical limit of $k \to \infty$, the $\theta$
coordinate complete decouples and becomes irrelevant.
The classical phase space coset of $(p_c, x_c)$ is
isomorphic to the space of pure states $(p^c, x^c)$ for
the observable algebra, and have the configuration space 
$x_c$ and momentum space $p_c$ as independent parts.
One can start with the coset $\tilde{G}(3)/ISO(3)\times T$
with $ISO(3)$ being the subgroup generated by $J_{ij}$ 
and $X_i$. We have
\bea \label{sc}
\left(\begin{array}{c}
dx^i  \\  d\theta \\    0
\end{array}\right) =
\left(\begin{array}{ccc}
 \omega^i_j &   0  &  \bar{x}^i \\
\bar{p}_{j}  & 0   &   \bar{\theta} \\
 {0}   & {0} &  0 
\end{array}\right)
\left(\begin{array}{c}
x^j \\ \theta \\    1
\end{array}\right)
\eea
the contraction limit of which also decouples $\theta$.
The coset space can be taken to give `coherent states'
of $\left|x^i \rra$ which are really position eigenstates.
The Hilbert space is however unitary equivalent to 
$\mathcal{H}$ above. The Hilbert space picture, or that 
of the Schr\"odinger wavefunctions $\phi(x^i)$ can
actually be used for which one can see that the
momentum operator $-i\hbar \partial_i$ goes to
the zero limit as $x$ is replaced by $x^c$ in the 
contraction.  The projective Hilbert space of the
reduced Hilbert space, or the space of pure states,
then becomes again isomorphic to the classical coset
of $x_c$. The picture cannot be used to look at the
Newtonian dynamics which cannot be described on
Newtonian space and functions on it alone. 


\section{DYNAMICS IN THE WWGM FORMALISM}
The most transparent mathematical formulation of the full
dynamical theory from the relativity symmetry perspective
is a Weyl-Wigner-Groenewold-Moyal (WWGM) formalism
 \cite{W,H,GV,D,Z} formalism with wavefunctions on the coherent 
state basis as the starting point. In fact, the representation on
$\mathcal{K}$ is really an irreducible component of the regular
representation of $H(3)$ and the operators $\hat{X}^{\ssc L}_i$
and $\hat{P}^{\ssc L}_i$ the reduction of the left invariant vector
fields. They can be written as $x_i\star$ and $p_i\star$,
respectively, with the $\star$ of the Moyal star product given by
\bea
\za \star \zb (p,x) = \za(p,x) e^{-i (\cev{\partial}_p \vec{\partial}_x-\cev{\partial}_x \vec{\partial}_p) } \zb(p,x) \;.
\eea
$\za(p,x)\!\star=\za(p\star,x\star)$ as a differential operator 
is exactly the representation of elements of the group 
$C^*$-algebra on the Hilbert space $\mathcal{K}$. The formalism
hence unifies the usual WWGM picture with the Hilbert space
one. Observables are functions of the position and momentum
operators $x_i\star$ and $p_i\star$ (Weyl ordering assumed), 
acting on the wavefunctions $\phi(p,x)$. An operator product
$\za\!\star\zb\star$ can be taken as $(\za\star\zb)\star$ which
is the notion of the Moyal star product. The Wigner distribution
for a state is given by the projection operator $\rho_\phi\star$ 
and explicitly $\rho_\phi(p,x) (\star) 
  = 2^{2n} \bar\phi(p,x)\star\phi(p,x) (\star)$. 
For a coherent state in particular, $\rho_{\phi_a}$ is a
real Gaussian centered at the expectation values, as expected.
Moreover, we have
\bea
\mbox{Tr} [\za\!\star \rho_\phi] 
=\frac{1}{2^2n}\frac{1}{\pi^n}  \int\!\! dp dx \; \za(p,x) \, \rho_\phi(p,x)
=\frac{1}{\pi^n}  \int\!\! dp dx \; \bar{\phi}(p,x) [ \za(p,x)\!\star {\phi}(p,x)] \;.
\eea

We have worked through the WWGM formalism in details 
to obtain the explicit results \cite{070} summarized above. 
However, one can really simply think about the whole story 
as writing down the full representation, again as a generic 
irreducible component of the regular representation of 
$H(3)$, for the group and the group $C^*$-algebra.
Moreover, the regular representation can be seen as 
essentially the quasi-regular representation of the full 
relativity symmetry on the coset $(p,x,\theta)$ as given
for the case of $\tilde{G}(3)$ above. It works as well for
the extended group of $H_{\!\ssc GH}(3)$. That picture is the
generic mathematical scheme that should work essentially
for any Lie group with the right representation.

With the contraction, we have in terms of $p^c$ and $x^c$
\bea \label{starc}
\star^c \sim
\exp \bigg[ \frac{-i\hbar}{k^2} (\cev{\partial}_{p^c}  \vec{\partial}_{x^c}  
    -\cev{\partial}_{x^c} \vec{\partial}_{p^c} ) \bigg] \to 1 \;,
\eea
{\em i.e.} the operator product reduces to 
$\za(p^c,x^c)\zb(p^c,x^c)$, namely the simple commutative
product of functions as $\za(p,x)\!\star$ becomes $\za(p^c,x^c)$.
The latter is the multiplicative operator on the Koopman-von
Neumann Hilbert space formalism particularly useful for describing
mixed states. A parallel picture at the quantum level can also be
formulated with the notion of a Tomita representation \cite{TT,trep}, 
which sees a density matrix (for a mixed state) as a vector in a 
Hilbert space (of operators). All that sits naturally in the functional
algebra $C(p,x)$.

The general symmetry on  ${\mathcal{K}}$ is described by  
the group of unitary transformations factored by its closed center 
of phase transformations. In particular, each one-parameter
subgroup of the relativity symmetry transformations is realized
as a (star-)unitary transformation in terms of real parameter $s$ 
as ${U}_{\star}\!(s)\star=e^{\frac{-i s}{2}{G}_{\!s}\star}$
with ${G}_{\!s}\star$ as the generator (the $2$ is $\hbar$).
${G}_{\!s}(p,x)$ is real as  ${G}_{\!s}\star$ is Hermitian.
For time translation, as a unitary transformation on 
${\mathcal{K}}$, we have the Schr\"odinger equation of motion 
\bea
2i \frac{d}{dt} \phi = {G}_{t}\!\star \phi \;.
\eea
The Heisenberg picture, for a generic ${G}_{\!s}\star$, gives
\bea\label{H}
 \frac{d}{ds} \za\star = \frac{1}{2i} [\za\star, {G}_{\!s}\star] \;.
\eea
Dropping the last $\star$, the equation is equivalent to
$\frac{d}{ds} \za = \frac{1}{2i} \{\za,  {G}_{\!s}\}_{\star}
=  \frac{1}{2i} [ \za\star{G}_{\!s} -{G}_{\!s}\star \za ]$
where $\{\cdot,  \cdot \}_{\star}$ is the Moyal bracket.
${G}_{t}\star$ is the Hamiltonian operator.

Under the contraction to the classical limit, the  
Schr\"odinger equation actually fails to give a proper
limit. That is actually not surprising. Recall that the quantum
Hilbert space for the pure states reduces to simple sum of
one dimensional subspace of the coherent states which
becomes essentially delta functions. Wavefunctions
$\phi(p^c,x^c)$ other than the delta functions no longer
describe pure states. In the picture of $\phi(p^c,x^c)$,
the pure states are simply no continuous distributed.
One can go to the Tomita representation which includes
the mixed states and look at the Liouville equation as
equation of motion for the density operator $\rho\star$,
which for the special case of a pure state is equivalent
to the Schr\"odinger equation. That has a proper limit as
\bea
 \frac{d}{dt} \rho(p^c,x^c;t) =\frac{k^2}{2i\hbar} \{ G_t^c(p^c,x^c), \rho(p^c,x^c;t)  \}_{\star^c} 
\to \{G_t^c(p^c,x^c), \rho(p^c,x^c;t) \}\;,
\eea
where  $\{\cdot, \cdot\}$ is  classical Poisson bracket 
$\{\za, \zb\}= \sum_i \left[ \frac{\partial\za}{\partial x^c_i}\frac{\partial\zb}{\partial p^c_i}
  - \frac{\partial\za}{\partial p^c_i}\frac{\partial\zb}{\partial x^c_i} \right]$.
Here, $G_t^c(p^c,x^c)$ is the classical Hamiltonian function.
Similarly, the Heisenberg equation of motion has the right 
classical limit as
 \bea \label{Heomk}
\frac{d}{dt} \za(p^c,x^c;t) = \frac{k^2}{2i\hbar}\{\za(p^c,x^c;t), G_t^c(p^c,x^c) \}_{\star^c} 
\to \{ \za(p^c,x^c;t), G_t^c(p^c,x^c) \}\;.
\eea
Moreover, all that works for any generic Hamiltonian flow
with generator $G_{\!s}$; Equation~(\ref{H}) gives the 
corresponding automorphism flow on the observable
algebra the classical limit of which is given in the form of
the Poisson bracket expression above.

We can define a
\bea \label{ug}
 \tilde{U}_\star\za\star=\mu(\za)\star=\bar{U}_\star\! \star \za  \star  {U}_\star\!\star \;,
\eea
and see $\tilde{U}_\star$ as a (star-)unitary operator
on some Hilbert space of operators. The Tomita
representation corresponds exactly to such a picture,
which gives on a density operator $\rho$ as a vector on the 
Hilbert space a `Schr\"odinger equation' in the form
\bea \label{leq}
\frac{d}{ds}\rho(s)   = \frac{1}{2i} \tilde{G}_{\!s} \rho(s)
 =\frac{1}{2i} \{{G}_{\!s},  \rho(s)\}_\star \;,
\eea
which is the Liouville equation. Furthermore, we can
actually look at ${G}_{\!s}$ and $\tilde{G}_{\!s}$ as
independent symmetry generators and hence get a
kind of doubled representation picture. For example,
we have for the position and momentum operators
\bea &&
G_{\!-x^i}\star = p_i\star\;, \qquad\qquad \tilde{p}_i = \tilde{G}_{\!-x^i} =2i \partial_{x^i} \;,
\sea
G_{\!p^i}\star = x_i\star \;, \qquad\qquad  \;\; \tilde{x}_i =\tilde{G}_{\!p^i} = 2i \partial_{p^i} \;.
\eea
Similar fundamental set of operators was long ago 
introduced within the Koopman-von Neumann 
formulation \cite{L}. The particularly interesting thing
is the following classical limits
\bea
\tilde{G}^c_{\!p^c} = i\hbar \partial_{p^c} \;,
\qquad\qquad
\tilde{G}^c_{\!-x^c} = i\hbar \partial_{x^c} \;,
\eea
giving translations in $p^c$ and $x^c$.  So, on the
Koopman-von Neumann Hilbert space for classical
mechanics, ${G}^c_{\!p^c}$ and  ${G}^c_{\!-x^c}$ as well
as all ${G}^c_{\!s^c}$ ($s^c$ may or may not equal to $s$)
are diagonal while generators of the type $\tilde{G}^c_{\!s^c}$ 
are not. It is the latter class of generators that
implement the classical relativity symmetries as
like translations and rotations. For the Galilean (free
particle) Hamiltonian, we have ($t^c=t$)
\bea
\tilde{G}_{t}^c 
= \frac{-i\hbar}{m}p_i^c\partial_{x_i^c} \;,
\eea
which match exactly to the classical equation of motion
as given in terms of the Poisson bracket.

\section{A NEW PICTURE FOR QUANTUM MECHANICS}
As a theory of particle dynamics on the quantum physical space,
quantum mechanics can be seen in a different light which has
a physical picture no less intuitive than classical mechanics.
In a way, that fully resolves issue in the famous Einstein-Bohr
debate which can be seen as paradoxical or counter-intuitive
because they have been formulated on concepts of the position
in space and other physical quantities described in terms of
a mathematical model which only works well as the classical
approximation. Using the right model, based first of all on
the quantum model of the physical space and extended to 
all physical quantities as in the observable algebra, we have
a successful abstract/mathematical formulation of physics
at least consistent with common sense intuitive notion 
provided that the latter is strip of any familiar but otherwise
abstract, hence not really physically intuitive, content from 
earlier versions of physical theory like the Newtonian
mechanics.

The Newtonian space with its three dimensional Euclidean
geometry, or for that matter any classical model of space(time)
modeled by a finite dimensional commutative geometry is only 
part of a classical theory, which fails at the quantum level. Such
commutative geometry is geometry modeled locally on products 
of the real number line. It has coordinate observables which can 
only be part of a commutative algebra. Taking the number of
dimension to the infinite limit, the picture become nontrivial.
However, there is really nothing intuitive about such finite
dimensional nature of the physical space. That is to say, 
there is nothing intuitive in the statement that the position
of a physical object , or the smallest indivisible part of one,
can be specified with exact precision by three (or a finite
number of) real numbers. The real number system is an
algebraic system developed to model the notion of a 
continuum. It is not even the only possible model of the 
latter though. Physicists have been assuming all physical
quantities are real-number valued. But that remains nothing
more than an assumption, or a model that works well for
classical physics. The real number answer is all practical
measurements come from the real number scale/reading
we ourselves put into the measuring apparatus in the
calibration. And all such reading has uncertainties. 
 
With quantum mechanics, we see that the observable algebra
is noncommutative. It can be seen as a functional algebra
$\za(\hat{P}_i,\hat{X}_i)$ of the three position and three
momentum observables the first model of which are 
operators on the quantum Hilbert space. In the meantime,
there has been much development in the mathematical
theory of noncommutative geometry \cite{C} essentially
seeing the position and momentum operators as the
noncommutative coordinates, here at least for the
noncommutative phase space. The projective Hilbert
space $\mathcal{CP}^\infty$ as the quantum phase space
is the space of pure states dual to the observable algebra.
Hence, it can be seen as an alternative description of
the noncommutative phase space in the language of 
real number geometry with infinite dimension. In fact,
it is easy to appreciate that the full physical content of
any operator can be described, under a choice of 
coordinate system, by infinite number of real numbers.
A naive example of the latter may be given by the set 
of matrix elements  on a chosen orthonormal basis of
the Hilbert space.  We have now illustrated that the
phase space is the right model for the physical space
behind quantum mechanics. It is also interesting to
note that the full observable algebra can also be
formulated as an algebra of (so-called K\"ahlerian)
functions on $\mathcal{CP}^\infty$ \cite{CMP}. What
is missing is a more physical intuitive description of
the geometry in terms of the noncommutative 
coordinates, on which we are now working.

The  $\mathcal{CP}^\infty$ is a K\"ahler manifold
hence a symplectic space. The Hilbert space itself is
also a K\"ahler manifold. The Schr\"odinger equation
is really a system of Hamiltonian equations for the
in terms of coordinates on the K\"ahler manifold. 
In terms of the natural complex coordinates on the
Hilbert space, given by $z^n=\lla z_n |\phi \rra$ for
an orthonormal basis $\left|z_n \rra$,  the real
parts serves as configuration variables and the 
imaginary parts the conjugate momentum variables;
the Hamiltonian function is 
$\frac{1}{2\hbar} \lla \phi|\hat{H} | \phi \rra$
taken as a function of $(p^n,q^n)$ with 
$z^n=q^n+ip^n$. Here, one can easily see that the
phase rotation 
$\left|\phi\rra \to e^{i\theta} \left|\phi\rra$ as
a basic (relativity) symmetry is a rotation on each
of the $q^n-p^n$ plane, showing clearly the 
division between configuration or position type 
and momentum type of coordinates is only a choice
of frame of reference. That is exactly like the
nature of the space and time coordinates of the 
Minkowski spacetime in Einstein special relativity. 

In the quantum/noncommutative physical space picture,
a quantum particle has a definite position given by a point
in the space, the solid mathematical description of which
require an infinite number of real coordinates or six
noncommutative coordinates. Such coordinates, in
principle, can be determined with arbitrary finite precision.
Such a particle also have fixed theoretical values for all 
observables without uncertainty. The value of each
observable, however, has to be described by an element
of a noncommutative algebra, or equivalently be an
infinite set of real numbers. We should think about that
as a kind of  noncommutative number, which encodes
a piece of quantum information about the particle. 
Uncertainties as in the Heisenberg uncertainty principle 
apply only to the best single real number description
of the value of an observable on a state, that is the
expectation value. In fact, in von Neumann, {\em i.e.} 
eigenvalue-answered, measurements, we need a good 
statistics from repeated measurements to get any
good approximation to the expectation value. And 
with good enough statistics, we can obtain also the 
uncertainty as the standard derivation as well as the
higher moments. The whole distribution, again a set
of infinite real numbers, is what truly characterizes 
the value or the full information of the observable 
on the state. Better/optimal descriptions of all that
is under study. 

A word about the notion of Born probability is in
order. The probability notion is quantum mechanics
obviously is about the real number modeling of 
physical position observables and other quantities.
Its implication on von Neumann measurements is only 
a statement about the statistical distribution. The 
latter physical result has also been illustrated as  
consequence of the decoherence induced by the 
measuring process. Adhering to the real number 
notion of all physical quantities and classical 
geometry for the physical space, the Born 
probability is the only workable way to interpret
a quantum state in relation to the possible or 
allowable observable values. With the quantum
physical space and noncommutative value of
observables, there is no probability notion 
necessary. God does not play dices, only that
he does not limit his mathematical tool used 
to build the physical universe to finite number 
of real numbers either.

\section{FURTHERING CONCLUDING REMARKS}
The notion of spacetime is the most fundamental
of all physical concepts.  Whatever fundamental
theory we have, it would have some basic
degrees of freedom which can be essentially
thought of as a description of the spacetime,
though such a spacetime model may be very 
different from the more familiar one. There would
also be a notion of relativity symmetry as the 
background symmetry of the theory and the
spacetime model, or the symmetry of reference
frame transformations. We can thing of a special
relativity for a kind of static background spacetime,
and a general relativity for the fully spacetime.
In this sense, a theory of quantum gravity would
be more like a theory of geometrodynamics for
a quantum spacetime, rather than some quantized
dynamical theory on a classical spacetime.

One can certainly think of  a quantum relativity
symmetry beyond the usual group theory framework,
such as using quantum groups. However, we see
from our studies that Lie groups are amazingly 
powerful. A noncommutative Lie group has
representations which extend to the group 
$C^*$-algebra are noncommutative. That looks
like is enough to describe a noncommutative
geometric space with any finite number of
noncommutative coordinate observables.
Formulating theory of the kind, with fully
noncommutative $\hat{X}_\mu$ and 
$\hat{P}_\mu$ is already quite a challenge. 
However, it is also a very promising one from
the relativity deformation/contraction 
perspective as sketched here. Quantum
mechanics and its classical approximation has 
been formulated with the kind of relativity 
symmetry perspective very successfully, 
as we reported here. The mathematical
scheme can be applied to the different Lie
groups, hence solidly anchoring such
theories to be developed on quantum
mechanics and Newtonian mechanics.

The Planck constant characterizing the 
noncommutativity in quantum mechanics
actually characterizes the constant holomorphic
sectional curvature of the projective Hilbert 
space \cite{CMP}. So, noncommutativity is
curvature, which then suggests the fascinating
idea of dynamical geometry implying dynamical
noncommutativity. 

Quantum field theory gives a very interesting
spacetime perspective. Seen properly, all the
quantum fields are more like degrees of freedom
for the spacetime itself. Finding the quantum
spacetime model incorporating that is
another important challenge ahead.

Back to quantum mechanics, we have discussed how
the theory can be seen in the light of noncommutative
value for observables. Full practical implementation
of the line of ideas may be like a Copernicus revolution
to the physics of measurements. A measurement is
just a controlled physical process to extract information
about a physical system. The quantum world has
information which is basically quantum. May be we can 
learn to extract and manipulate quantum information
as kind of noncommutative numbers. There may be
the days when physicists will calibrate our apparatus
with noncommutative numbers.

\section{ACKNOWLEDGMENTS}
The author is partially supported by research grants 
number 107-2119-M-008-011 of the MOST of Taiwan.

\nocite{*}
\bibliographystyle{aipnum-cp}%

\end{document}